\documentstyle[12pt]{article}
\begin{document}

\begin{titlepage}

\hfill{September 1995}

\hfill{}

\hfill{UM-P-95/90}

\hfill{RCHEP-95/21}

\vskip 1 cm

\centerline{{\large \bf
Large neutrino asymmetries from neutrino oscillations
}}

\vskip 1.5 cm

\centerline{R. Foot, M. J. Thomson and R. R. Volkas}

\vskip 1.0 cm
\noindent
\centerline{{\it Research Centre for High Energy Physics,}}
\centerline{{\it School of Physics, University of Melbourne,}}
\centerline{{\it Parkville, 3052 Australia. }}

\vskip 1.0cm

\centerline{Abstract}
\vskip 1cm
\noindent
We re-examine neutrino oscillations in the early universe.
Contrary to previous studies, we show that large neutrino
asymmetries can arise due to oscillations between
ordinary neutrinos and sterile neutrinos. This means
that the Big Bang Nucleosynthesis (BBN)
bounds on the mass and mixing of ordinary
neutrinos with sterile neutrinos can be evaded. Also,
it is possible that the neutrino asymmetries can be large
(i.e. $\stackrel{>}{\sim} 10\%$), and hence
have a significant effect on BBN through nuclear reaction
rates.

\end{titlepage}

There are several experimental indications that
neutrinos have nonzero mass and oscillate\cite{snd, ana,
lsnd}. It is possible that sterile neutrinos exist
which oscillate with the known neutrinos.
There are essentially two types of sterile
neutrinos that can be envisaged.
First, there are sterile states which either
have no gauge interactions, or interactions which are
much weaker than the usual weak interactions \cite{sn}.
Alternatively, it is possible to envisage neutrinos which
do not have significant interactions with ordinary matter
but do have significant interactions with themselves.
An interesting example of the latter is
given by mirror neutrinos\cite{flv}.

However, for both sterile and mirror neutrinos
there are apparently quite stringent bounds
if they are required to be
consistent with standard big bang cosmology.
Assuming that the effective number of
neutrino species present during Big Bang Nucleosynthesis (BBN)
is bounded to be
less than 3.4, then the mixing angle ($\theta_0$) and the squared
mass difference ($\delta m^2$) for a sterile neutrino mixing with
one of the known neutrinos is approximately
bounded by \cite{B}
$$ |\delta m^2| (\sin^2 2\theta_0)^{1.8}
\  \stackrel{<}{ \sim}\
10^{-8} \ eV^2, \  \nu=\nu_{e},$$
$$ |\delta m^2| (\sin^2 2\theta_0)^{1.6}
\  \stackrel{<}{ \sim}\
10^{-7} \ eV^2, \  \nu=\nu_{\mu,\tau}, \eqno (1)$$
assuming that $\delta m^2 < 0$ and
$\sin^2 2\theta_0 \ \stackrel{<}{\sim}\ 10^{-3}$ (for
more precise bounds see e.g. figure 4 of Ref.\cite{dm}).
These bounds arise by demanding that oscillations
not bring the sterile neutrino into equilibrium with the
known neutrinos.
These bounds (along with even more stringent bounds
for the case $\delta m^2 > 0$ or $\sin^2 2\theta_0 \approx 1$)
would appear to exclude the region of parameter
space required to explain the atmospheric neutrino anomaly
in terms of $\nu_{\mu}-\nu_s$ oscillation ($|\delta m^2| \simeq 10^{-2}
\ eV^2$, $\sin^2 2\theta_0 \simeq 1$),
and would restrict the parameter space required to explain
the solar neutrino deficit in terms of $\nu_e - \nu_s$ oscillation.
An important assumption in deriving the bounds of Eq.(1) was that
the relic neutrino asymmetries could be neglected.
The purpose of this paper is to re-examine this issue.
We will show that for a significant region of
parameter space large neutrino asymmetries can be
generated due to neutrino oscillations.
This has two important consequences. First,
the bounds in Eq.(1) will be modified. Second,
the asymmetries can be large enough to affect
the neutron/proton ratio and hence also modify BBN.
This is one plausible way
of reconciling the possible disagreement of the BBN
predictions with observations\cite{hata}.

To define our notation, we
 first examine the ordinary neutrino ($\nu_{\alpha}$,
$\alpha = e, \mu, \tau$) oscillating
with a sterile neutrino ($\nu_s$) in vacuum. Oscillations can occur
if the weak eigenstate
neutrino and sterile neutrino are each
linear combinations
$$\nu_{\alpha} = \cos\theta_0 \nu_1 + \sin\theta_0 \nu_2, $$
$$\nu_s = -\sin\theta_0 \nu_1 + \cos \theta_0 \nu_2, \eqno (2)$$
of mass
eigenstates $\nu_{1,2}$.
An ordinary neutrino of momentum $p$
will then oscillate in vacuum after a time $t$
with probability
$$|\langle \nu_{\alpha} (t)|\nu_s\rangle|^2 = \sin^2 2\theta_0
\sin^2 \left({t \over L_{osc}}\right), \eqno (3)$$
where \cite{c2}
$$L_{osc} = {2p \over \delta m^2} \equiv {1 \over \Delta_0}.\eqno (4)$$
However, in the early universe oscillations occur in a plasma.
For $\nu_{\alpha} - \nu_s$
oscillations in a plasma of temperature $T$, the matter and vacuum
oscillation parameters are related by \cite{msw}
$$\sin^2 2\theta_m = {\sin^2 2\theta_0 \over 1 - 2z \cos 2\theta_0
+ z^2},$$
$$\Delta_m^2 = \Delta_0^2 ( 1 - 2 z \cos 2\theta_0 + z^2 ),
\eqno (5)$$
where $z = 2\langle p \rangle
\langle V_{\alpha} - V_s \rangle /\delta m^2$
and $\langle V_{\alpha, s} \rangle$ are the effective potentials
due to the
interactions of the neutrinos with matter ($\langle p \rangle \simeq 3.15
T$). For a truly sterile neutrino $V_s = 0$.
For a weak eigenstate
neutrino $\nu_{\alpha}$, $V_{\alpha}$ is given
by \cite{nr, ekm2, c}
$$V_{\alpha} = \sqrt{2} G_F N_{\gamma}\left( L^{(\alpha)} -
{A_{\alpha} T^2 \over M_W^2}\right), \eqno (6)$$
where $G_F$ is the Fermi coupling constant, $M_{W}$
is the $W$ boson mass, $A_{\alpha}$ is a numerical factor
given by $A_e \simeq 55$ and $A_{\mu, \tau} \simeq 15.3$ \cite{nr,ekm2}.
The quantity $L^{(\alpha)}$ is given by
$$L^{(\alpha)} = L_{\alpha} + L_{\nu_e} + L_{\nu_{\mu}} +
L_{\nu_{\tau}} + (\pm {1 \over 2} + 2x_w)L_e +
({1 \over 2} - 2x_w)L_p - {1 \over 2}L_n,
\eqno(7)$$
where the plus sign refers to $\alpha = e$ and the minus sign
refers to $\alpha = \mu, \tau$. Also, $x_w \equiv \sin^2 \theta_w \simeq
0.23$ and
$L_{\alpha} \equiv (n_{\alpha} - n_{\bar \alpha})/n_{\gamma}$.

We now study the possible creation of lepton number due to
oscillations. This issue has been studied previously \cite{ekm2, bd}
where it was concluded that significant neutrino asymmetries
could not be generated due to neutrino oscillations.
That conclusion was based on studies covering a
limited region of parameter space (in particular
for very small values of $|\delta m^2| \stackrel{<}{\sim} 10^{-5}
eV^2$).
We will show below that for a large range of parameters
it is possible to generate large neutrino asymmetries. Note
that in Ref.\cite{bd} it was observed that the neutrino
asymmetry could experience a brief period of exponential growth
for appropriate parameters. However, the particular parameters
they chose led to a final asymmetry of only about $10^{-7}$.
The analysis below will demonstrate that much larger asymmetries
(up to $10^{-1}$ and even larger)
can be generated for other interesting parameter
choices.

We will work analytically, and then check our work numerically
using the density matrix formalism.
Our analytic treatment assumes that the change in
lepton number is dominated by collisions, which is true
for temperatures greater
than a few MeV\cite{fn}. We will also assume that the evolution
of the neutrino ensemble follows the evolution of the state
with average momentum.

The change in lepton number due to collisions can be
expressed in terms of reaction rates as follows\cite{fn}:
$$\begin{array}{ccc}
{\delta (n_{\nu_{\alpha}} - n_{\bar  \nu_{\alpha}})\over \delta t}
& \simeq & - n_{\nu_{\alpha}}\Gamma (\nu_{\alpha}
\to \nu_s) + n_{\nu_s}\Gamma(\nu_s \to \nu_{\alpha}) \\
& + &
n_{\bar \nu_{\alpha}}\Gamma(\bar \nu_{\alpha} \to \bar \nu_s
) - n_{\bar \nu_s}\Gamma(\bar \nu_s
\to \bar \nu_{\alpha})
\end{array}. \eqno(8)$$
If the number densities of ordinary neutrinos
are equal to the number densities of
sterile neutrinos, then
$\delta L_{\nu_{\alpha}}/\delta t \to 0$.
This is because the rate $\Gamma(\nu_{\alpha} \to \nu_s)$ equals
$\Gamma(\nu_s \to \nu_{\alpha})$ (and similarly for the antiparticle
rates). However, if the number of sterile neutrinos is
negligible to start with, and if they
are not brought into equilibrium
with the ordinary neutrinos, then $\delta L_{\nu_{\alpha}}/\delta t$
is nonzero in general (provided $\delta m^2 <0, \cos2\theta_0 > 0$,
as we shall demonstrate).
We thus consider parameters satisfying the bounds
Eq.(1), so that the sterile neutrinos are not brought into
equilibrium with the ordinary neutrinos.

Thus, assuming $n_{\nu_s}, n_{\bar \nu_s} \ll n_{\nu_{\alpha}},
n_{\bar \nu_{\alpha}}$,
the change in lepton number due to collisions can be expressed as,
$$\frac{\delta L_{\nu_{\alpha}}}{\delta t} \simeq
{3 \over 8}\left[ - \Gamma(\nu_{\alpha} \to \nu_s)
+ \Gamma
(\bar \nu_{\alpha} \to \bar \nu_s)\right]
- {L_{\nu_{\alpha}} \over 2}\left[\Gamma(\nu_{\alpha} \to
\nu_s) + \Gamma(\bar \nu_{\alpha} \to \bar
\nu_s)\right],  \eqno (9)$$
where we have used $n_{\nu_{\alpha}} = 3n_{\gamma}/8$
[recall $L_{\nu_{\alpha}}
\equiv (n_{\nu_{\alpha}} - n_{\overline{\nu}_{\alpha}})/n_{\gamma}$]. It
will turn out that the second term (proportional to
$L_{\nu_{\alpha}}$)
can be neglected for $T \stackrel{>}{\sim} 1 MeV$.

Thus, in order to evaluate
$\delta L_{\nu_{\alpha}}/\delta t$, we need to
evaluate the reaction rates.
The rate $\Gamma(\nu_{\alpha} \to \nu_s)$ is given by
the interaction rate of ordinary neutrinos  multiplied
by the probability that the neutrino collapses to the
sterile eigenfunction, i.e.
$$\Gamma(\nu_{\alpha} \to \nu_s) =
\langle P_{\nu_{\alpha} \to \nu_s}\rangle_{coll} \Gamma_{\nu_{\alpha}}
\eqno (10)$$
where $\Gamma_{\nu_{\alpha}} = y_{\alpha} G_F^2 T^5 $(with $y_e \simeq 4.0,
y_{\mu, \tau} \simeq 2.9$)
are the reaction
rates\cite{B}. The quantity
$\langle P_{\nu_{\alpha} \to \nu_s}\rangle_{coll}$ is the
probability that the neutrino $\nu_{\alpha}$ collapses to the
sterile state $\nu_s$. It is given by
$$\langle P_{\nu_{\alpha} \to \nu_s}\rangle_{coll}
= \sin^2 2\theta_m \langle \sin^2 {x \over L_{osc}^{(m)}}
\rangle, \eqno (11)$$
where $x$ is the distance between collisions.
Note that
$\langle x \rangle \equiv L_{int} = 1/\Gamma_{\nu_{\alpha}}$
where $L_{int}$ is the mean distance between interactions.
In the region where $L_{int} \gg L^{(m)}_{osc} (\bar L^{(m)}_{osc})$,
$\langle\sin^2 x/L^{(m)}_{osc}\rangle \to 1/2$ and
$\langle\sin^2 x/\bar L^{(m)}_{osc}\rangle \to 1/2$ ($L^{(m)}_{osc}$
and $\bar L^{(m)}_{osc}$ being the oscillation lengths for neutrinos
and antineutrinos, respectively).
Thus, evaluating $\delta L_{\nu_{\alpha}}/\delta t$ in the region where
$L_{int} \gg L^{(m)}_{osc}(\bar L^{(m)}_{osc})$, we find that\cite{fn2}
$${\delta L_{\nu_{\alpha}} \over \delta t} \simeq
{3 \over 16}\Gamma_{\nu_{\alpha}}
\left( - \sin^2 2\theta_{m_{\alpha}}
+ \sin^2 2\theta_{m_{\overline{\alpha}}} \right).
\eqno (12)$$
Recall that $\sin^2 2\theta_{m_{\alpha}, m_{\overline{\alpha}}}$
are defined in Eq.(5), where $z$ is given by
$$z =  -a + b,\ {\rm for}\ \nu = \nu_{\alpha}\quad {\rm and}$$
$$z = a + b,\ {\rm for}\ \nu = \bar \nu_{\alpha}, \eqno (13)$$
where
$$ a \equiv -\sqrt{2}G_F n_{\gamma}L^{(\alpha)}/\Delta_0,\
b \equiv  -\sqrt{2}G_F n_{\gamma} A_{\alpha}T^2/(\Delta_0 M_W^2).
\eqno (14)$$
Equation (12) then becomes
$${\delta L_{\nu_{\alpha}} \over \delta t} =
{3 \Gamma_{\nu_{\alpha}} \sin^2 2\theta_0
a(\cos2\theta_0 - b)\over
4[1 - 2\cos2\theta_0 (-a+b) + (a-b)^2][1 - 2\cos2\theta_0
( a+b) + (a + b)^2]}. \eqno (15)$$
This contribution to $\delta L_{\nu_{\alpha}}/\delta t$ is
larger than the term which we neglected [the second term of
Eq.(9)] provided that $a/(1 + a^2)
\gg 4L_{\nu_{\alpha}}/3\cos2\theta_0$, i.e.
$T \stackrel{>}{\sim} 0.4(|\delta m^2|/\cos2\theta_0
eV^2)^{1/4}$.
Thus, for $T \stackrel{>}{\sim}\ 1 MeV$ the
second term of Eq.(9) can be approximately neglected and
Eq.(15) should be
correct. Note that for temperatures of order
1 MeV and below, the effect of the second term in Eq.(9) is to reduce
$|L_{\nu_{\alpha}}|$.

We will assume that $\delta m^2 < 0$ (and hence $a, b > 0$) and
$\cos2\theta_0 > 0$\cite{fn3}.
Observe that for $b > \cos2 \theta_0$,
$L^{(\alpha)} = 0$ is a stable fixed point.
To see this, note that
if $L^{(\alpha)} >0$ then
$\delta L^{(\alpha)}/\delta t <0$ while for $L^{(\alpha)} < 0$,
$\delta L^{(\alpha)}/\delta t >0$.
The critical observation is that when $b < \cos 2\theta_0$,
$L^{(\alpha)} = 0$ becomes an unstable fixed point (i.e. if
$L^{(\alpha)} > 0$ then $\delta L^{(\alpha)}/\delta t > 0$ while
for $L^{(\alpha)} <0$, $\delta L^{(\alpha)}/\delta t < 0$).
Since $b$ is proportional to $T^6$, at some point
during the evolution of the universe $b$ becomes
less than $\cos 2\theta_0$ and
$L^{(\alpha)} = 0$ becomes unstable.
Observe that for $a \ll \cos2\theta_0$,
$\delta L_{\nu_{\alpha}}/\delta t$ is
proportional to $L_{\nu_{\alpha}}$
leading to a rapid exponential growth of $L_{\nu_{\alpha}}$.

Calculating the temperature when $b = \cos 2\theta_0$, we find
$$T_c \simeq 12.9 (16.2)  \left({\cos 2\theta_0 |\delta m^2|
\over eV^2}\right)^{1/6} MeV,
\eqno (16)$$
for electron (muon/tau) neutrinos.

In order to calculate the amount of $L_{\nu_{\alpha}}$
produced, it turns out
that it is more convenient to use the variable $a$.
This is because Eq.(15) has simple forms depending on whether
$a \ll 1$ or $a \gg 1$.
The quantity $\delta a/\delta t$, unlike
$\delta L_{\nu_{\alpha}}/\delta t$,
has a contribution from the expansion
of the universe.
Calculating $\delta a/\delta t$, we find,
$${\delta a \over \delta t} =
{\partial a \over \partial t}|_{exp} + {\partial a \over \partial
t}|_{osc} =
{\partial T \over \partial t}{\partial a \over \partial T}
+ {\partial a \over \partial L_{\nu_{\alpha}}}{\partial
L_{\nu_{\alpha}} \over \partial t} $$
$$ \simeq {-5.5 T^3 \over M_P}{4a \over T} -
{2\sqrt{2}G_F n_{\gamma} \over \Delta_0}{\delta L_{\nu_{\alpha}}\over
\delta t},
\eqno (17)$$
where we have used the time-temperature relation
$t \simeq M_P/11 T^2$ which is relevent for low temperatures (i.e.
$2m_e < T < m_{\mu}$). We have also used the result
that $a$ is proportional to $T^4$ (so that $\partial a /\partial T
= 4a/T$) and that $a$ is linear
in $L_{\nu_{\alpha}}$. Thus, using Eq.(15) we find that
$${\delta a \over \delta t} \simeq {21T^2a \over M_P} \left[
{\lambda T^7 (\cos 2\theta_0 - b)
\over [(\cos2\theta_0 +a-b)^2 + \sin^2 2\theta_0]
[(\cos2\theta_0 -a-b)^2 + \sin^2 2\theta_0]}
\  - \ 1\right],
\eqno (18)$$
where $\lambda$ is a constant given by
$$\lambda \simeq {-0.11y_{\alpha}G_F^3 \sin^2 2\theta_0 M_P \over
 \delta m^2}. \eqno (19)$$
Note that $\lambda > 0$ since we are assuming that $\delta m^2 < 0$.
Observe also that $\delta a/\delta t > 0$ for $0 < a < a_c$ (where
$a_c$ is a number greater than $\cos2\theta_0$ which we will define
precisely below),
provided that $T^7 \cos 2\theta_0
 > 1/\lambda$ for $b \ll \cos2\theta_0$ (for $b \sim \cos2\theta_0$,
i.e. at resonance, the above behaviour
holds for even smaller values of $T$).
For $a\gg \cos2\theta_0$, $\delta a/\delta t < 0$.
This means that the parameter $a$
evolves towards a non-zero value:
$a \to a_c \stackrel{>}{\sim} \cos2\theta_0$.
To calculate $a_c$ we solve the equation
$${\delta a  \over \delta t} = 0, \eqno (20)$$
to find that
$$a_c^2 = {K + \sqrt{K^2 - 4C} \over 2},
\eqno (21)$$
where
$$K = 2(\cos 2\theta_0 - b)^2 - 2\sin^2 2\theta_0, $$
$$C = (1-2\cos2\theta_0 b + b^2)^2 -
\lambda T^7 (\cos 2\theta_0 - b). \eqno (22)$$
Recall that $L_{\nu_{\alpha}}$ is related to $a$ through
Eq.(14). Hence, the result that $a$ evolves to  $a_c$ is
equivalent to the statement that $L_{\nu_{\alpha}} $ evolves to
a non-zero value $L^c_{\nu_{\alpha}}$,
where $L^c_{\nu_{\alpha}}$ is given by
$$L^c_{\nu_{\alpha}} = {\Delta_0 a_c \over 2\sqrt{2}G_Fn_{\gamma}}.
\eqno (23)$$
To summarise the situation: $L_{\nu_{\alpha}}$ evolves to
a very small value (so that $L^{(\alpha)} \to 0$) until
the temperature is such that $b=\cos2\theta_0$. At this point
$L^{(\alpha)} = 0$ becomes an unstable fixed point and
$L_{\nu_{\alpha}}$
exponentially increases until some point $a_c \stackrel{>}{\sim}
\cos2\theta_0$. Strictly this behaviour will not occur for
all values of $\delta m^2, \sin^2 2\theta_0$. For $\sin^2 2\theta_0$
small enough, the evolution of $a$ will be such that it will
never reach the value $a_c$. Nevertheless,
there is a large range of parameters for which $a$
evolves to $a_c$.

For later times, $a_c$ remains greater than
$\cos2\theta_0$ until
$${\delta a|_{exp} \over \delta t} > max \left({\delta a|_{osc}\over
\delta t}\right).  \eqno (24)$$
The maximum value of $\delta a|_{osc}/\delta t$ occurs at the
resonance i.e. $a \simeq \cos 2\theta_0$.
Solving the above equation
we find that\cite{fn4}
$$T_x \simeq \left({9(1+3\cos^2 2\theta_0) |\delta m^2| \over
M_P y_{\alpha} G_F^3\cos2\theta_0}\right)^{1/7}. \eqno (25)$$
For $\cos2\theta_0 \sim 1$ then
$$T_x \simeq \left({(|\delta m^2|/eV^2)
\over 2}\right)^{1/7} \ MeV. \eqno (26)$$
Thus $a_c \ge \cos2\theta_0 \sim 1$ for $T \ge T_x$. For later
times ($T < T_x$), $\delta a|_{exp}/\delta t >
max(\delta a|_{osc}/\delta t)$
and $L_{\nu_{\alpha}}$ is essentially frozen.
The relation $a_c \simeq \cos2\theta_0$ and $T = T_x$ implies
that the corresponding value for $L_{\nu_{\alpha}}$
will be
$$L_{\nu_{\alpha}}
\simeq 2 \times 10^{-2} (|\delta m^2|/eV^2)(T_x)^{-4}. \eqno (27)$$
Recall that our analysis is really only valid for
temperatures greater than about 1 MeV. For $|\delta m^2| < 30 \ eV^2$
we should take $T_x$ to be of order $1 MeV$.

 From Eq.(27) it is clear that we expect that large
neutrino asymmetries can be generated.
In fact, for $|\delta m^2| \stackrel{>}{\sim} 5 \ eV^2$,
$L_{\nu_{\alpha}} \stackrel{>}{\sim} 0.1$.
We have checked our results numerically using the
density matrix formalism
(see e.g.\cite{mt} for a description of this formalism
and for original references).
Numerically integrating the equations describing the evolution of
the density matrix we have found good agreement with the analytic
results presented in this paper. In figure 1 we plot the evolution
of $L_{\nu_{\alpha}}$ taking, by way of example,
$\delta m^2 = -1 \ eV^2, \sin^2 2\theta_0 \simeq
10^{-8}$. Also plotted in figure 1 is the asymptotic value
$L^c_{\nu_{e}}$ [Eq.(23)].
Figure 1 shows the behaviour expected from our analytic
analysis. The lepton number $L_{\nu_e}$ initially grows exponentially
at the resonance $b = \cos 2\theta_0$ [which corresponds
to 12.9 MeV for $\delta m^2 = -1 \ eV^2$, $\sin^2 2\theta_0 =
10^{-8}$ according to Eq.(16)].
After the exponential phase gets cut off,
the neutrino asymmetry continues to grow more slowly as it approaches
the asymptotic curve determined from $a_c$.

One important result
is that the lepton number generated can be large enough to
significantly modify the standard BBN scenario. Indeed
if $L_{\nu_e}$ is positive \cite{fluc} then it can reduce
the neutron to proton ratio. (Such a reduction may
be one way  to get agreement with the data \cite{hata}).

Another important result of this work is that the bounds
on ordinary neutrinos mixing with sterile neutrinos may
be aleviated. We will study this issue in more
detail and expand our analysis in a
forthcoming article using the density matrix formalism.
Below we sketch how this can happen.
By way of example, assume that there is a sterile
neutrino which
mixes with the muon neutrino, with parameters suggested
by the atmospheric neutrino anomaly (i.e. $|\delta m^2| \simeq
10^{-2} eV^2$, $\sin ^2 2\theta_0 \simeq 1$\cite{ana}).
Naively, this possibility is in conflict with BBN,
because these parameters severely violate the BBN bounds of
Eq.(1).
However a small mixing between the tau neutrinos and the
sterile neutrino may generate significant
lepton number to strongly suppress the oscillations
of the muon and sterile neutrino \cite{fv}.
This would then make this scenario consistent with
BBN.

Consider a system comprised of $\nu_{\mu}$, $\nu_{\tau}$ and $\nu_s$.
Note that for oscillations of tau to sterile neutrinos,
the mass difference can be much bigger than the $10^{-2} eV^2$ for
muon to sterile neutrinos.
[$\delta m^2_{\tau s} =  m_{\nu_s}^2 -
m^2_{\nu_{\tau}} \simeq -m_{\nu_{\tau}}^2$].
Denote $b_{\tau} (b_{\mu})$ as the value of the $b$ parameter
[defined in Eq.(14)] with $\delta m^2 = \delta m^2_{\tau s}
(\delta m^2_{\mu s})$.
Because of the larger mass difference and the larger value
of $\cos 2\theta_{\tau s}$, the point $b_{\tau} = \cos2\theta_{\tau s}$
will be reached
earlier for $\nu_{\tau}-\nu_s$ oscillations
than  the corresponding point $b_{\mu} = \cos 2\theta_{\mu s}$
for $\nu_{\mu} - \nu_s$ oscillations.
If the vacuum parameters for $\nu_{\tau} -\nu_s$
mixing, $\theta_{\tau s}, \delta m^2_{\tau  s}$ satisfy
the BBN bounds of Eq.(1), then $\nu_s$ cannot be brought into
equilibrium by $\nu_{\tau} -\nu_s$ oscillation.
Consequently our earlier analysis applies and we would
expect significant generation of tau lepton number at this point.
Note that the creation of $L_{\nu_{\tau}}$
could conceivably be compensated
by $\nu_{\mu} \to \nu_s$ oscillations.  We
assume that there is significant
intergenerational mixing between $\nu_{\mu}$ and $\nu_{\tau}$
which we assume  rapidly
converts $L_{\nu_{\tau}}$ into
$L_{\nu_{\mu}}$ so that
$L_{\nu_{\mu}} \simeq L_{\nu_{\tau}}$. This assumption
may not really be necessary (we will study this in more
detail in our forthcoming article\cite{ftv}). The condition
that the creation of lepton number due to $\nu_{\tau} - \nu_s$
oscillations will not be erased by $\nu_{\mu} -\nu_s$
oscillations is
$$|{\delta L_{\nu_{\tau}} \over \delta t}| >
|{\delta L_{\nu_{\mu}} \over \delta t}|,
\eqno (28)$$
for temperatures satisfing $b_{\tau} < \cos2\theta_{\tau s}$.
Studying the above inequality
away from resonance by using Eq.(15)
(with $b_{\mu}/b_{\tau} = \delta m^2_{\tau  s}/\delta m^2_{\mu s}$),
we find that it is sufficient to demand that
$$\sin^2 2\theta_{\tau s} > \sin^2 2\theta_{\mu s} \left({\delta m^2_{\mu s}
\over \delta m^2_{\tau s}}\right)^2.
\eqno (29)$$
If $\sin^2 2\theta_{\mu s} \approx 1$ and $\delta m^2_{\mu s} \approx 10^{-2}
eV^2$ then
$$\sin^2 2\theta_{\tau s} > 10^{-4} {eV^4 \over m^4_{\nu_{\tau}}}.
\eqno (30)$$
Thus, we arrive at a lower bound for $\sin 2\theta_{\tau s}$,
which for $m_{\nu_{\tau}} < 30 eV$ (as suggested by cosmology)
is given by $\sin^2 2 \theta_{\tau s} \stackrel{>}{\sim} 10^{-10}$.

We also need to check that at the point where the lepton number
begins to be generated [i.e. when $b_{\tau} = \cos2\theta_{\tau s},$
see Eq.(16)], the parameters
are such that the $\nu_s$ have
not already been brought into equilibrium.
Demanding that ${1 \over 2}\Gamma_{\nu_{\alpha}}\sin^22\theta_m <
H \simeq 5.5 T^2/M_P$ for $T > T_c$ [defined in Eq.(16)], we find:
$${|\delta m^2_{\mu s}| \over |\delta m^2_{\tau s}|}\
\stackrel{<}{\sim}
\ {\cos 2\theta_{\tau s}
\over 41[\cos2\theta_{\tau s} |\delta m^2_{\tau s}|]^{1\over 4}
\sin2\theta_{\mu s}}.
\eqno (31)$$
So, for $|\delta m^2_{\mu s}| \simeq 10^{-2}\ eV^2,
\cos2\theta_{\tau s} \simeq 1$ and $ \sin2\theta_{\mu s}
\simeq 1,$ we get that
$m_{\nu_{\tau}} \stackrel{>}{\sim}\ 0.5\  eV$.

Finally we need to check that the lepton number generated
is sufficient to suppress the $\nu_{\mu} - \nu_s$ oscillations.
Demanding that interactions not bring the sterile neutrino
into equilibrium with the muon neutrino, i.e.
$\Gamma_{\nu_s} \stackrel{<}{\sim} H \simeq 5.5T^2/M_P$,
we find in the case of large $L^{(\mu)}$\cite{fv} that
$$(\delta m_{\mu s}^2)^2 \stackrel{<}{\sim}
{79G_F^2 T^2 n_{\gamma}^2 [L^{(\mu)}]^2 \over
{y_{\alpha}M_P G_F^2 T^3 \over 11} - 1},\eqno (32)$$
where $T > T_{dec} \simeq 4.4 MeV$ (since we only need to
require that the sterile neutrinos not come into equilibrium
before kinetic decoupling of the muon neutrinos occcur).
Since from our earlier analysis, we expect
$a_c \stackrel{>}{\sim} \cos2\theta_{\tau s}$,
we know that the $L^{(\tau )}$ generated by the $\nu_{\tau}-\nu_s$
oscillations satisfies
$$|L^{(\tau)}| \stackrel{>}{\sim}
{|\delta m^2_{\tau s}| \over 2\sqrt{2} G_F n_{\gamma} 6.3T}.
\eqno (33)$$
Thus assuming that intergenerational mixing rapidly
distributes the lepton number so that $L^{(e)} \simeq L^{(\mu)}
\simeq L^{(\tau)}$ then Eq.(32) is satisfied
provided that
$$|\delta m_{\tau s}^2| \stackrel{>}{\sim} 10 |\delta m_{\mu s}^2|,
\eqno(34) $$
where the most stringent bound occurs when at the decoupling
temperature i.e. $T = T_{dec} \simeq 4.4 MeV$.
In other words, for the parameters satisfying the above equation
there is sufficient lepton number generated to prevent
the sterile neutrino from coming into equilibrium (above
$T_{dec}$) with
the muon neutrino [despite having oscillation parameters in
gross violation of the bounds Eq.(1)].
Assuming $\delta m_{\tau s}^2 \simeq -m_{\nu_{\tau}}^2$
and $|\delta m_{\mu s}^2| \simeq 10^{-2} \ eV^2$
(as suggested by the atmospheric neutrino
anomaly) then Eq.(34) suggests
that $m_{\nu_{\tau}} \stackrel{>}{\sim} 0.3 \ eV$.
This requirement is slightly less stringent then the $0.5 \ eV$
bound obtained earlier from Eq.(31).
Thus we conclude that muon neutrino oscillating into
a sterile neutrino with maximal mixing and $\delta m^2
\simeq 10^{-2} \ eV^2$ as suggested by the
atmospheric neutrino anomaly seems to
be consistent with BBN provided that
the tau neutrino oscillates into the sterile neutrinos
with parameters bounded by
$10^{-4}/m^4_{\nu_{\tau}} < \sin^2 2\theta_{\tau s}
\stackrel{<}{\sim}
3 \times  10^{-5}/m_{\nu_{\tau}}^{1.2}$ [where we have used
Eqs.(1) and (30)]. Note that this  condition implies that
$m_{\nu_{\tau}} > 1.5 \ eV$.

This result should also apply for mirror neutrinos  which do not
have significant interactions with ordinary particles but do
have significant interactions with themselves\cite{flv}.
Mirror neutrinos are required to exist if there
is an unbroken parity symmetry.
However, naively this scenario appears to be in conflict
with BBN because it predicts that the muon neutrino
will be maximally mixed
with a mirror neutrino. Thus it will violate the bounds Eq.(1)
if it is to be relevant for the atmospheric neutrino anomaly.
However, we have shown in this paper that this scenario
is not in conflict with BBN provided that the sterile neutrino
also mixes slightly with the tau neutrino.

Thus we conclude that it is possible that large neutrino
asymmetries can be generated by oscillations between
ordinary and sterile neutrinos. This asymmetry can be very
large (i.e. $\stackrel{>}{\sim} 0.10$) and may thus
modify nucleosynthesis. This should provide one way
to reconcile the present discrepancy between the
BBN predictions and observations
\cite{hata}. Another consequence of a large neutrino
asymmetry is that it may be possible to aleviate the BBN
bounds on the mixing of ordinary and sterile neutrinos.
\vskip 1.3cm
\noindent
\centerline{\bf Acknowledgements}
\vskip .3cm
This work was supported by the Australian Research Council.

\vskip  1.3cm
\centerline{\bf Figure Caption}
\noindent
Electron lepton number ($L_{\nu_e}$) versus temperature
(in MeV units) for $\nu_e - \nu_s$ oscillations with parameters
$\delta m^2 = -1 \ eV^2$ and $\sin^2 2\theta_0 = 10^{-8}$.
The broken line is the theoretical prediction based
on the numerical integration of the density matrix
equations. The solid line is the asymptotic value
$L^c_{\nu_e}$ [Eq.(23)].


\begin{thebibliography}{99}
\bibitem{c2}
Actually $|L_{osc}|$
is the oscillation length divided by $2\pi$.
\bibitem{c}
For antineutrinos, the equation is the same as below, except that
$L^{(\alpha)} \rightarrow -L^{(\alpha)}$.
\bibitem{fn}
We can also calculate the change in lepton number due to
oscillations. For $L_{int} \gg L_{osc}$ (where $
\langle\sin^2 x/L_{osc}\rangle \to 1/2$), the number density
of neutrinos and antineutrinos is given by
$n_{\nu_{\alpha}} = 1 - {\sin^2 2\theta_{m_{\alpha}}\over 2},
\ n_{\bar \nu_{\alpha}} = 1 - {\sin^2 2\theta_{m_{\bar \alpha}} \over 2}$
(where we are not including the depletion due to collisions).
Hence,
$${\partial(n_{\nu_{\alpha}} - n_{\bar \nu_{\alpha}}) \over \partial t}
= {\partial \over \partial t}\left(
{\sin^2 2\theta_{m_{\bar \alpha}}\over 2}
- {\sin^2 2\theta_{m_{\alpha}}\over 2} \right).$$
This equation has a similiar form to the contribution to
$\delta L_{\nu_{\alpha}}/\delta t$ due to collisions [Eq.(15)],
but it is numerically smaller provided that $y_{\alpha}G_F^2 T^5
> 8T^2/M_P$ i.e. for $T \stackrel{>}{\sim} 2 MeV$.
\bibitem{fn2}
Note that it is possible that $L_{int}/L_{osc} < 1$ at
the center of the resonance.
Indeed, $$\left({L_{int} \over L_{osc} }\right)^2 =
{\Delta^2_0[(\cos2\theta_0 - (b\pm a))^2 + \sin^2 2\theta_0]
\over (y_{\alpha} G_F^2 T^5)^2}$$
$$    \approx {(b \pm a - \cos 2\theta_0)^2 + \sin^2 2\theta_0
\over (b^2/A_{\alpha}^2)}$$
where we have used $y_{\alpha} G_F^2 T^5/\Delta_0 \approx b/A_{\alpha}$.
However, if
$|b \pm a - \cos 2\theta_0 |/b \gg 1/A_{\alpha},$
then $L_{int}/L_{osc} \gg 1$, and hence our replacement
of $\langle\sin^2 x/L_{osc}\rangle $ by $1/2$ should be valid
in the resonance region, but it may not be
strictly valid at the center of the resonance.
\bibitem{fn3}
Note that if we had taken $\delta m^2 > 0$, then $a, b < 0$ and
$L^{(\alpha)} = 0$ would be a stable fixed point for all values of $a, b$
provided that $\cos2\theta_0 > 0$. For $\cos 2\theta_0 < 0$
we require $\delta m^2 > 0$ for $L^{(\alpha)} = 0$ to be an unstable fixed
point.
\bibitem{fn4}
Note that strictly, at resonance we may be in a region
where $L_{int}/L_{osc} < 1$ and consequently Eq.(15) is
not appropriate.  However recalculating $max
(\delta a|_{osc}/\delta t)$ in the limit
where $L_{int} \ll L_{osc}$ (at the resonance) we
find that $\delta a|_{exp}/\delta t$ is always less
than $\delta a|_{osc}/\delta t$.
Eventually,
$L_{int} > L_{osc}$ at resonance. Thus,
$T_x$ is given by the above, or when $L_{int}/L_{osc} \approx 1$,
which ever is later.
\bibitem{fluc}
Note that we cannot predict the sign of $L^{(\alpha)}$. For
temperatures where $b > \cos2\theta_0$, $L^{(\alpha)}$ evolves
very near to zero. So near to zero that for temperatures where
$b < \cos 2\theta_0$, it is likely that
the sign of $L^{(\alpha)}$ will depend on random statistical
fluctuations.
It is amusing to speculate that if $L_{\nu_e}$ evolves to a large value
so that the ratio of protons to neutrons is affected, then it
may be possible that different regions of
the early universe have different signs
of $L_{\nu_e}$ and thus the ratio of protons and neutrons may not
be homogeneous today!
\end{thebibliography}
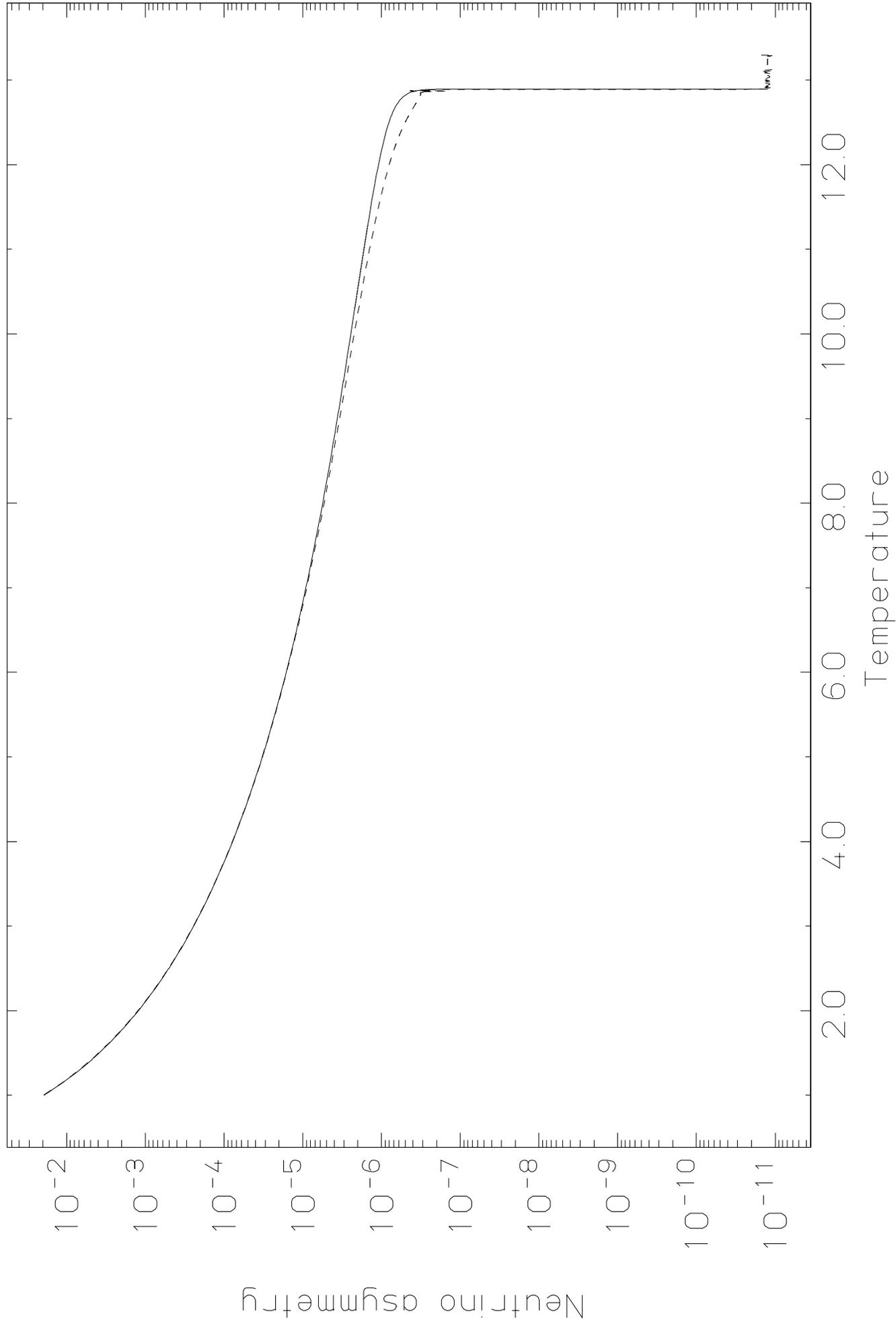
\end{document}